\documentclass[11pt,a4]{article}

\RequirePackage{fancybox}
\RequirePackage{amsmath}
\RequirePackage[dvips]{graphicx}
\RequirePackage{amssymb}

\RequirePackage[font=footnotesize,labelfont=bf]{caption}
\RequirePackage{subcaption}
\RequirePackage{float}

\RequirePackage[T1]{fontenc}

\RequirePackage{comment}

\RequirePackage{algpseudocode,algorithm}

\RequirePackage{hyperref} 

\usepackage[misc]{ifsym}

\usepackage{tikz}
\usetikzlibrary{shapes, arrows}
\usetikzlibrary{positioning}

\DeclareGraphicsExtensions{ps}
\graphicspath{{fig/}{fig1/}{tmp/}{../fig/}}
\makeatletter
  \newcommand\figcaption{\def\@captype{figure}\caption}
  \newcommand\tabcaption{\def\@captype{table}\caption}
\makeatother

\def\mi{\begin{equation}}
\def\mf{\end{equation}}

\def\mia#1\mfa{\begin{align}#1\end{align}}
\def\miar#1\mfar{\begin{eqnarray}#1\end{eqnarray}}
\def\mmi#1\mmf{\begin{multline}#1\end{multline}}
\def\<#1\>{\begin{equation}#1\end{equation}}
\def\[#1\]{\begin{equation}#1\end{equation}} 

\def\reff#1{(\ref{#1})}

\newcommand{\ud}{\mathrm{d}}

\renewcommand{\v}[1]{\ensuremath{\mathbf{#1}}} 
\newcommand{\gv}[1]{\ensuremath{\mbox{\boldmath$ #1 $}}}  
\let\baraccent=\= 
\renewcommand{\=}[1]{\stackrel{#1}{=}} 

\newcommand*{\plot}[2][]{\noindent\includegraphics[width=\ifx &#1& 0.98\else #1\fi\linewidth]{#2}\vskip -0.8cm\figcaption{}} 

\RequirePackage{color}
\RequirePackage[normalem]{ulem}  

\newif\ifprintcallout
\printcallouttrue

\title{Transmission matrix parameter estimation of COVID-19 evolution with age compartments using ensemble-based data assimilation}
\author{Santiago Rosa$^{\text{\Letter}}$ (1, 2), Manuel Pulido (2, 3), Juan Ruiz (4, 5), Tadeo Cocucci (1, 2)}
\date{}
\graphicspath{{./figuras/}}

\begin{document}

\maketitle

\begin{center}
    \Letter\hspace{.2cm} santiago.rosa@mi.unc.edu.ar
\end{center}

\bigskip
\noindent (1) FaMAF, Universidad Nacional de C\'ordoba, C\'ordoba, C\'ordoba, Argentina
\\
(2) FaCENA, Universidad Nacional del Nordeste, Corrientes, Corrientes, Argentina
\\
(3) CNRS - IRD - CONICET - UBA, Instituto Franco-Argentino para el Estudio del Clima y sus
Impactos (IRL 3351 IFAECI), Buenos Aires, Argentina
\\
(4) CONICET - Universidad de Buenos Aires, Centro de Investigaciones del Mar y la Atmósfera
(CIMA), Buenos Aires, Argentina
\\
(5) Departamento de Ciencias de la Atmósfera y los Océanos, FCEN, Universidad de Buenos Aires, Ciudad Aut\'onoma de Buenos Aires, Buenos Aires, Argentina

\bigskip

\begin{abstract}
  The  COVID-19 pandemic and its multiple outbreaks have challenged governments around the world. Much of the epidemiological modeling was based on pre-pandemic contact information of the population, which changed drastically due to governmental health measures, so called non-pharmaceutical interventions made to reduce transmission of the virus, like social distancing and complete lockdown. In this work, we evaluate an ensemble-based data assimilation framework applied to a meta-population model to infer the transmission of the disease between different population agegroups. We perform a set of idealized twin-experiments to investigate the performance of different possible parameterizations of the transmission matrix. These experiments show that it is not possible to unambiguously estimate all the independent parameters of the transmission matrix. However, under certain parameterizations, the transmission matrix in an age-compartmental model can be estimated. These estimated parameters lead to an increase of forecast accuracy in agegroups compartments assimilating age-dependent accumulated cases and deaths observed in Argentina compared to a single-compartment model, and reliable estimations of the effective reproduction number. The age-dependent data assimilation and forecasting of virus transmission may be important for an accurate prediction and diagnosis of health care demand.
\end{abstract}

\section{Introduction}

Governments around the world have had to make several difficult  decisions with the widespread of the SARS-COV-2 virus in early 2020. Different flavors of social distancing measures from localized risk population to general lockdowns were implemented to alleviate the propagation of COVID-19, at the expense of a decline in the productivity. A lockdown may have a strong impact on the epidemic propagation with a flattening of the active cases curve. On the other hand, it also has a negative impact in the education and social activities. Furthermore, a large COVID-19 outbreak also affects the economy, as evidenced in the case of widespread and strictly enforced sick leaves. Therefore, decision makers need to evaluate carefully the trade-off between socio-economical well-being and sanitary conditions. There is a need to develop real time decision making tools which can monitor the situation of the pandemic and be able to predict the evolution of the disease at different scales: from neighborhoods and cities to states and nationwide. The epidemiological predictions may help to prevent some overloading of the health system: different analysis of thresholds and tendencies of the amount of active cases may be used by governments to implement different non-pharmaceutical interventions which can prevent the collapse of healthcare availability. Research on COVID-19 spread monitoring and modelling (e.g.\cite{ferguson20}) had a strong political impact worldwide: several governments around the globe opted for various actions after it. However, despair COVID-19 evolution in different countries made clear that a continuous monitoring of the local spreads based on data was required to adopt timely distancing measures. This work is the result of a project from a grant call for COVID-19 research of the Research National Agency in Argentina in which real time prediction of the propagation, and in particular the epidemic peaks, around the country was one of the main objectives.

COVID-19 propagation has been modeled through epidemiological models, most commonly population compartmental models, like Susceptible-Exposed-Infected-Recovered (SEIR) models. In some cases, these models may give good estimations, particularly at the initial phase of an outbreak. However, the virus propagation is subject to the complexity of human interactions or individual-wise varying viral loads \cite{grossman21}, which are difficult to describe with compartmental models. Even the most advanced meta-population models (e.g. GLEAM \cite{balcan09}) and agent based models \cite{kerr20} represent very crudely the transmission dynamics of the virus since it depends on said interactions between individuals which are difficult to model and (most importantly) predict in a realistic fashion. Furthermore, social life  changed significantly through the evolution of the pandemic because of several factors (government decisions, news, social status).

The accumulated data about the epidemic is also rather limited and prone to errors: detection policies have changed with time, delay in reported cases occurring on weekends, lack of hospital discharges dates, etc. On top of the mentioned sources of uncertainty, there is a large amount of undetected cases: a large number of individuals does not suffer noticeable symptoms and/or they do not report them, or, in a smaller scale, the tests give false negatives \cite{lau21}. Since data are incomplete and noisy and models suffer from misrepresentation of the underlying complex processes, the idea of combining model and data becomes appealing. The main aim of real-time model-data fusion techniques, referred to as sequential inference or data assimilation, is to combine very diverse sources of information considering their uncertainties. In particular, data and epidemiological models are considered with their uncertainty and the techniques aim to: determine the epidemiological state of the population, estimate the optimal model parameters and  quantify the optimal model uncertainty represented via stochastic processes, using the observational evidence.

One of the most advanced techniques for prediction and risk assessment are those associated with weather forecast events implemented in environment prediction centers and national weather services. These agencies need to model climate disasters including flash floods, extreme droughts and heat waves. There is a plethora of observational instruments of the atmosphere such as satellites, airplanes, radiosondes, and meteorological radars. Data is being generated continuously by these instruments and need to be fuse with numerical model predictions. Furthermore, there are substantial regions on Earth which are poorly observed (e.g. vast areas over southern oceans). State of the art data assimilation methods combining numerical models and data are essential to propagate information between different variables, both spatially and temporally for weather forecasting\cite{carrassi18}.  This process is conducted in real-time. There is a standardized protocol for the meteorological data acquisition and storage, and modelling for an optimal communication and collaboration between countries and/or state agencies. \cite{Schemm23} propose to organize similar international protocols for  epidemiological modeling.

There are some works that apply data assimilation techniques for epidemiological modeling. Shaman et al \cite{shaman12}, \cite{shaman13} use an ensemble-based data assimilation framework to model influenza propagation. The state evolution of an epidemiological model, i.e. SIRS model, is combined with direct and indirect data (e.g. level of web activity related with the illness) from the epidemic. At the same time,  parameters of the system are learned online as the observations become available. In that work, they use a variant of the ensemble Kalman filter (EnKF). An EnKF estimation and forecast cases of Cholera applied to a SIRB model (the B stands for the concentration of \textit{V. cholerae} in water reservoirs) divided into communities is conducted in \cite{pasetto17}. The model is forced with the amount of rainfall each community experienced, and assimilating weekly cases and deaths, it allows to forecast new cases.

With the necessity of monitoring the spread of COVID-19 and because of the worldwide abundance of data, several works used data assimilation to estimate the spread of the SARS-COV-2 virus.
Li et al. \cite{li20} use the iterated filter-ensemble adjustment Kalman filter to assimilate COVID-19 data within China using a meta-population model and mobility data. They propose the estimation of the undocumented (asymptomatic) infections fraction together with  the rate of transmission of the undocumented infections. They estimate the undocumented rate to be 86\%.
Engbert et al. \cite{engbert21} use an EnKF for regional transmission modeling. They propose maximizing the likelihood to estimate time-independent parameters in a stochastic SEIR model to capture the dynamic of the epidemic at regional levels.
Evensen et al.  \cite{evensen21} applied an ensemble Kalman smoother technique  to a meta-population model. The evolution of epidemiological parameters is estimated over a long time period assuming a prior density for them. The technique is able to capture the abrupt change in the reproduction number in several countries after lockdown measures.

Chinazzi et al. \cite{chinazzi20} use a meta-population epidemiological model combining the individual spreads between regions, via flight information. The reproduction number ($R_0$) is estimated using approximate Bayesian computation varying $R_0$ and comparing the resulting simulations with the observed number of imported cases. 

There is a  strong dependence between the severity of COVID-19 symptoms and age. Infections among children and young people often result in asymptomatic cases. On the other hand, adults aged over 60 develop the most severe, and sometimes lethal, cases. Transmission effects have also been associated to age \cite{stringhini20}, \cite{dattner20}, \cite{davies20}. While children under 10 years old appear to have a low susceptibility to infection, people over 60 are highly susceptible. Identifying age-dependence in the virus transmission is essential for policy making using non-pharmaceutical interventions, e.g. school opening/closing \cite{evensen21}. 

Estimating the amount of contacts between individuals for a particular population is a challenge, and it is usually achieved by statistically significant population surveys. Klepac et al. \cite{klepac18} use the data collected from a smartphone application in the UK to infer social interactions: The data contains the contact history of each user labeled by agegroups, so the authors have an empirical statistical contact matrix of the population. This was used in an ABM to simulate an influenza-like outbreak for the BBC documentary \textit{Contagion}.
  Arregui et al. \cite{arregui18} use surveys from eight countries \cite{Mossong08} to extrapolate known contact matrices to other countries.
These works use a fixed contact matrix to study the evolution of epidemics and there is no estimation of time-varying contact rates. \cite{evensen21} use a base matrix $\mathcal{C}$ modulated by a time-dependent coefficient $R(t)$, in which case the transmission matrix is $\Lambda=R(t) \mathcal{C}$ . The base matrix is normalized in such a way that $R(t)$ is the effective reproduction number.

This work aims to study alternatives to time-independent transmission matrix, proposing time-varying parameterizations. Along these parameters, we also estimate relevant parameters, like the effective reproduction number and fraction of detected cases and deaths, using information about the age-structured data of the virus spread. The changes over time in the transmission matrix are also estimated (e.g. changes in mobility in one of the age groups considered). To this end, we combine a meta-population SEIRHD model with a stochastic EnKF to assimilate age-structured cumulative cases and deaths. Finally, we forecast the age-dependent propagation of COVID-19.

The outline of this article is as follows: In section \ref{metodos} we show our model and introduce the data assimilation framework. In section \ref{detalles_exp} we give details of the real-world data utilized, present the general experimental details and show the different contact matrix parameterizations used. In section \ref{resultados} we present and discuss the results, each subsection corresponds to a different experiment including synthetic and real-world data experiments. In section \ref{conclusiones} we draw the conclusions of our investigation.

\section{Technique details} \label{metodos}
\subsection{Compartmental epidemiological model} \label{modelo}

The evolution of  COVID-19 is modeled for the whole population of a region, which is assumed to be isolated. The model we used is an extension of a basic SEIR (Susceptible, Exposed, Infectious, Recovered) model, where a closed population (i.e. no births, deaths, immigration or emigration) is divided into $n$ agegroups. A detailed description of classic SEIR models may  be found in \cite{blackwood18}. The variables considered are $S_j$ (susceptible), $E_j$ (exposed but not infectious), $I_j$ (infected), $M_j$ (mild symptoms), $T_j$ (severe symptoms), $C_j$ (critical symptoms), $R_j$ (recovered) and $D_j$ (deaths). The index $j=1,\,...\,,n$ is used to indicate the agegroup.

\begin{figure}[H]
    \centering
    \includegraphics[width=.7\textwidth]{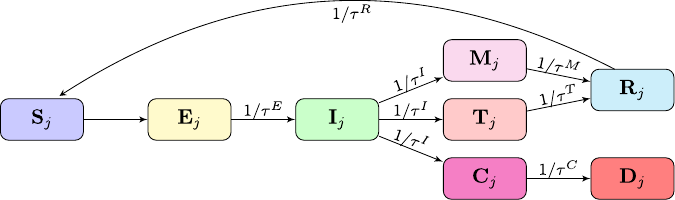}
    \caption{\footnotesize Diagram of the compartmental model. An individual moves to the next compartment after a period $\tau^{\mathcal{X}}$, which depends of the compartment $\mathcal{X}$.}
    \label{fig:flowchart1}
\end{figure}

The flow between epidemiological categories of the model is shown in Fig \ref{fig:flowchart1}. Infected individuals in the agegroup $j$, $I_j$, can interact with  susceptible  individuals in the agegroup $k$, $S_k$, with a transmission rate $\lambda_{jk}$. The individuals of the susceptible classes $S_j$ that are exposed to the disease are moved to the exposed compartment $E_j$. The individuals in this compartment do not transmit the virus. After a mean incubation time $\tau^E$, the exposed individuals move to the infected group $I_j$. In this stage, individuals can  spread the virus to susceptible persons during the period $\tau^I$. After that, the individuals transit to the compartments $T_j$, $C_j$ or $M_j$ with probabilities $f^T_j$, $f^C_j$ and $1-f^T_j-f^C_j$, respectively. The group $T_j$ contains the individuals presenting severe cases that require hospitalization and, after a time $\tau^T$, recovers from the disease moving to the recovered individuals compartment $R_j$. The compartment $C_j$ (critical) represents the individuals with severe cases that require hospitalizations and, after a time $\tau^C$, die and move to the dead compartment $D_j$. The compartment $M_j$ consists of the individuals who present mild symptoms and require no hospitalization, and after a time $\tau^M$, they transit to the recovered compartment. After a period $\tau^R$, individuals from the recovered compartment becomes susceptible again given that SARS-COV-2 immunity diminishes substantially after 5-7 months \cite{tyler20}.
The compartments are designed to characterize the COVID-19 infection dynamics. Individuals are unable to transmit the virus in the initial incubation phase and then  are infectious during a period. They are also expected to be isolated once the symptoms are apparent (or tested positive). Therefore, once the individuals transit to $M_j$, $T_j$ or $C_j$ they are expected to be isolated and do not spread the disease, only individuals in the compartment $I_j$ do.

The model parameters are the transmission matrix parameters $\lambda_{jk}$ (which is the number of contacts  that a  persons in group $j$ have with persons in group $k$, in a period of time $\Delta_t$, multiplied by the probability of a contact resulting in an infection), the average time an individual stays in each of the epidemiological states $\tau^E$, $\tau^I$, $\tau^M$, $\tau^T$, $\tau^C$ and $\tau^R$, and the fractions of infections $f_j^T$ and $f_j^C$ of $I_j$ moving to $T_j$ and $C_j$, respectively. The population of each age compartment is constrained by the total population of the age group.

The resulting model equations are

\begin{equation}
    \begin{split}
                                    N_j &= S_j + E_j + I_j + M_j + T_j + C_j + R_j + D_j\\
        \frac{\partial S_j}{\partial t} &= -\frac{S_j}{\tau^I N_j}\, \sum_{k=1}^n \lambda_{jk}\, I_k + \frac{R_j}{\tau^R}\\
        \frac{\partial E_j}{\partial t} &= \frac{S_j}{\tau^I N_j}\, \sum_{k=1}^n \lambda_{jk}\, I_k - \frac{E_j}{\tau^{E}}\\
        \frac{\partial I_j}{\partial t} &= \frac{E_j}{\tau^{E}}-\frac{I_j}{\tau^{I}}\\
        \frac{\partial T_j}{\partial t} &= f^T_j \frac{I_j}{\tau^I} - \frac{T_j}{\tau^{T}}\\
        \frac{\partial C_j}{\partial t} &= f^C_j \frac{I_j}{\tau^I} - \frac{C_j}{\tau^{C}}\\
        \frac{\partial M_j}{\partial t} &= (1-f^T_j-f^C_j) \frac{I_j}{\tau^I} - \frac{M_j}{\tau^{M}} \\
        \frac{\partial D_j}{\partial t} &= \frac{T_j}{\tau^{C}} \\
        \frac{\partial R_j}{\partial t} &= \frac{M_j}{\tau^{M}} + \frac{T_j}{\tau^{T}} - \frac{R_j}{\tau^R}
    \end{split}
    \label{eq:model}
\end{equation}

Table \ref{table1} summarizes the variables and the parameters and Table \ref{table2} shows the numeric values of all the fixed parameters except for the transmission matrix, which is one of the variables to be estimated.

\begin{table}[H]
    \begin{center}
    \footnotesize
    \begin{tabular}{|l | l|}
        \hline 
        \multicolumn{2}{|l|}{Variables}  \tabularnewline [1ex]
        \hline
        $S_j$ & Susceptible  individuals  \\
        $E_j$ & Exposed individuals (non-contagious yet)    \\
        $I_j$ & Infected individuals (contagious)    \\
        $M_j$ & Infected with mild symptoms (isolated)  \\
        $T_j$ & Individuals with severe symptoms that eventually will recovered (isolated) \\
        $C_j$ & Individuals with critical symptoms that eventually will die (isolated)  \\
        $R_j$ & Recovered   \\
        $D_j$ & Dead    \\
        \hline
        \multicolumn{2}{|l|}{Parameters}  \tabularnewline [1ex]
        \hline
        $\lambda_{jk}$ & Transmission rate between the agegroup k to j \\ 
        $\tau^E$ & Incubation period. \\
        $\tau^I$ & Infection period. \\
        $\tau^M$ & Recovery period for mild infections. \\
        $\tau^T$ & Recovery period for sever infections. \\ 
        $\tau^C$ & Time until death. \\
        $\tau^R$ & Time until immunity vanishes. \\
        $f^T_j$ & Fraction of the infected individuals in the agegroup $j$ that develops severe symptoms.\\   
        $f^C_j$ & Fraction of the infected individuals in the agegroup $j$ that eventually dies.\\
        $N_j$ & Total number of individuals in the  agegroup $j$.\\
        \hline
    \end{tabular}
    \end{center}
    \caption{Model variables and parameters.}
    \label{table1}
\end{table}

\begin{table}[H]
    \begin{center}
    \footnotesize
    \begin{tabular}{|c|c|c|c|}
        \hline
        \multicolumn{4}{|c|}{Agegroup-dependent parameters}\\
        \hline
        Agegroup    & 1 & 2 & 3  \\
        Age range   & 0-29 & 30-64 & 65-103 \\
        $f^T$ & 0.1 & 0.05 & 0.26 \\
        $f^C$ & 0.002 & 0.009 & 0.095\\
        \hline
        \multicolumn{4}{|c|}{global parameters}\\
        \hline
        $\tau^E$ & \multicolumn{3}{c|}{4} \\
        $\tau^I$ & \multicolumn{3}{c|}{5} \\
        $\tau^M$ & \multicolumn{3}{c|}{7} \\
        $\tau^T$ & \multicolumn{3}{c|}{15} \\ 
        $\tau^C$ & \multicolumn{3}{c|}{15} \\
        $\tau^R$ & \multicolumn{3}{c|}{150} \\
        \hline
        \end{tabular}
    \caption{Numeric value of the model parameters. All time scales ($\tau$'s) are expressed in days.  }
    \label{table2}
    \end{center}
\end{table}


The parameters controlling the propagation of a disease in a meta-population model are the transmission matrix elements. In a population divided in age-compartments, they represent the interaction between the infected and susceptible agegroups and hence it is the main driver of the disease evolution.  One of the central objectives of this work is to parameterize and estimate the transmission matrix to obtain a better representation of the propagation between agegroups. In Eq. \reff{eq:model}, the elements of the transition matrix are not independent \cite{arregui18}: the total amount of contacts, that   individuals of the group $j$ have with individuals of the group $k$, has to be equal to the total amount of contacts that individuals of the group $k$ have with individuals of the group $j$:

\begin{equation}
    \lambda_{jk}N_j=\lambda_{kj}N_k.
    \label{eq:trm_constraint}
\end{equation}


The most relevant parameter in epidemiological modeling is the basic reproduction number. It represents the mean number of new infected individuals caused by one infected person in a totally susceptible population. The  basic reproduction number may be estimated in compartmental models by linearising the dynamics of the infected differential subsystem, which is the part of the model that governs the production of new infections when all individuals are susceptible (in a SEIR model, for example, this subsystem are the compartments SEI). The resulting Jacobian matrix is known as \textit{next generation matrix} \cite{diekmann90}, whose spectral radius corresponds to the basic reproduction number $R_0$.  If the linearization of the infected subsystem  is conducted at time $t$, the spectral radius of the resulting matrix is known as the effective reproduction number $R_{\text{eff}}$. This represents the amount of secondary cases that an infected individual produces at time $t$, assuming the remaining of the non-infected or recovered population is susceptible.  A review of the topic can be found in \cite{heffernan05}.

Let us assume $m$ compartments $x_m$  in a compartmental epidemiological model whose individuals can transmit the disease and define the vector $\v x=(x_1,x_2,\,...\,x_m)$, which in our model is $(E_1,E_2,\,...\,,E_n, I_1, I_2\,...\,,I_n)$. If $F_i$ is defined as the rate of appearance of new exposed individuals and $V_i$ the balance of the entry and exit (by the natural progression of the disease) of individuals in the $i$-th compartment, then, the rate of change of a variable $x_i$ in the model is given by
  
\begin{equation}
    \frac{\partial x_i}{\partial t} = F_i(\v x) - V_i(\v x).
    \label{eq:rate_change_FV}
\end{equation}

Next, the Jacobian matrices of the dynamical system, $\mathbb{F}$ and $\mathbb{V}$, are defined as
\begin{equation}
    [\mathbb{F}]_{ij} = \frac{\partial F_i}{\partial x_j}, \,\,\,\,\,\,\,\,\,\,\, [\mathbb{V}]_{ij} = \frac{\partial V_i}{\partial x_j}.
    \label{eq:mat_FV}
\end{equation}

The next generation matrix  $\mathbb{G}$ is then defined as $\mathbb{G}=\mathbb{F}\mathbb{V}^{-1}$. The $ij$ element of $\mathbb{G}$ is interpreted as the rate at which infected individuals in $x_j$ produce new infections in $x_i$  times the average amount of time they spend in the compartment $j$. The effective reproduction number $R_{eff}$ is the largest absolute eigenvalue, i.e. the spectral radius of the next generation matrix. In this work, the next generation matrix and the resulting effective reproduction number are inferred with the data assimilation system as  diagnostic information of the epidemic.

\subsection{State-parameter estimation with ensemble-based data assimilation methods} \label{enkf}

The evolution of epidemiological variables can be modeled as a partially observed time evolving process, i.e. a hidden Markov model. Within this framework, the evolution of the state of the system can be written as

\begin{equation}
    \textbf{x}_{k+1} = \mathcal M(\textbf{x}_{k}) + \gv \eta_k,
    \label{eq:markov_model}
\end{equation}

\noindent where $\textbf{x}_k$ is the state of the system at time $k$, $\mathcal M()$ is the dynamical model and  $\gv \eta_k$ is the model error. The second equation forming the hidden Markov model corresponds to the observational map. The observations $\v y_k$ are related to the state $\textbf{x}_k$ by the observation operator ${\bf H}_k$ which maps the space of state variables to the observational space

\begin{equation}
    \textbf{y}_k = {\bf H}_k \textbf{x}_k + \gv \epsilon_k,
    \label{eq:observations}
\end{equation}

\noindent where $\gv \epsilon_k$ is the observation error. In this work, ${\bf H}_k$ is assumed linear but the method can be generalized to the non-linear case. In filtering theory, the estimation problem involves obtaining the conditional probability density function (pdf) of $\textbf{x}_k$ knowing the current and past observations $\v Y_k = ({\textbf{y}_1,\textbf{y}_{2},...,\textbf{y}_{k}})$, denoted by $p(\textbf{x}_{k}|\textbf{Y}_{k})$ (a.k.a. filtering or analysis distribution). We can obtain the prediction pdf by performing a forecast step

\begin{equation}
    p(\textbf{x}_k | \textbf{Y}_{k-1}) = \int d\textbf{x}_{k-1} \, p(\textbf{x}_{k}|\textbf{x}_{k-1}) \, p(\textbf{x}_{k-1}|\textbf{Y}_{k-1})
    \label{eq:forecast}
\end{equation}
then, using Bayes theorem, the posterior density conditioned on the set of observations is obtained:

\begin{equation}
    p(\textbf{x}_{k}|\textbf{Y}_k) = \frac{p(\textbf{y}_k|\textbf{x}_k) p(\textbf{x}_k|\textbf{Y}_{k-1})}{ \int d\textbf{x}_k p( \textbf{y}_k|\textbf{x}_k) p(\textbf{x}_k|\textbf{Y}_{k-1}) }.
    \label{eq:analisis}
\end{equation}

Eqs. \reff{eq:forecast} and \reff{eq:analisis} can be solved sequentially every time new observations $\textbf{y}_k$ are available, but they have to be integrated over the entire state space, which is usually computationally intractable. However, using a sample based representation of the distributions, the forecast step can be approximated by a Monte Carlo approach by simply evolving every sample point forward with the model $\mathcal M()$. In this work we use the EnKF, which is a Monte Carlo non-linear extension of the Kalman Filter \cite{kalman60}. The analysis distribution is represented by an ensemble of possible states. The resulting analysis state members are of the form

\begin{equation}
    \textbf{x}^{\text{a},(i)} = \textbf{x}^{\text{f},(i)} + \textbf{K}(\textbf{y}^{(i)} - \v H\textbf{x}^{\text{f},(i)})
    \label{eq:kalman_filtering}
\end{equation}

\begin{equation}
    \textbf{K} = \textbf{P}^\text{f}\textbf{H}^\top (\textbf{H} \textbf{P}^f \textbf{H}^\top + \textbf{R})^{-1}
    \label{eq:kalman_gain}
\end{equation}
\noindent where $\v R$ is the observational covariance matrix (assumed known), and the forecast error covariance $\textbf{P}^f$ is estimated from the ensemble of forecasted state vectors:

\begin{equation}
    \overline{\textbf{x}}^\text{f} = \frac{1}{m}\sum^m_{i=1} \textbf{x}^{\text{f},(i)} \, , \,\,\,\,\,\,\,\,\,\,\, \textbf{P}^\text{f} = \frac{1}{m-1}\sum^m_{i=1}(\textbf{x}^{\text{f},(i)} - \overline{\textbf{x}}^\text{f})(\textbf{x}^{\text{f},(i)} - \overline{\textbf{x}}^\text{f})^\top.
    \label{eq:cov_forecast}
\end{equation}

The analysis ensemble mean,

\begin{equation}
    \overline{\textbf{x}}^\text{a} = \frac{1}{m}\sum^m_{i=1} \textbf{x}^{\text{a},(i)},
\end{equation}

\noindent provides a point estimate of the state of the system.

In Eq. \reff{eq:kalman_filtering}, the observation vector is perturbed with Gaussian noise: $\textbf{y}^{(i)} = \textbf{y} + \gv\mu^{(i)}$, where $\gv\mu^{(i)} \sim N(\textbf{0},\textbf{R})$. This is required  to obtain a sample covariance of the analysis state members with the expected analysis covariance \cite{burguers98}.

During the analysis update, the EnKF can result in non-physical values for some model parameters and ensemble members (e.g. negative values for the transmission matrix elements). This is a consequence of the assumption of Gaussian forecast error in the EnKF. To avoid this complication, we force the lower limit of all the estimated parameters to $0$ in each ensemble member. 

The observation operator is linear, and the conversion from state space to observation space is as follows:  we assume that all variables except $S_j$ and $E_j$ are partially documented. This is achieved with a parameter $0<\gamma_j<1$ in the observational operator, ${\bf H}$, which accounts for the sub-detection of cases. In other words, we assume there is a sub-detection bias in some observational variables. This parameter depends on the agegroup, since the symptoms may increase with age, so that the amount of asymptomatic cases is larger for children. In the agegroup $j$, the relation between the cumulative observed cases ($y^c_j$) and observed deaths ($y^d_j$) and the state variables at time $k$ is

\begin{equation}
    \begin{pmatrix}
        y^c_j\\
        y^d_j
    \end{pmatrix}
    =
    \begin{pmatrix}
        0 & 0 & \gamma_j    & \gamma_j  & \gamma_j  & \gamma_j  & \gamma_j    & \gamma_j \\
        0 & 0 & 0           & 0         & 0         & 0         & 0             & 1
    \end{pmatrix}
    \begin{pmatrix}
        S_j\\
        E_j\\
        I_j\\
        M_j\\
        T_j\\
        C_j\\
        R_j\\
        D_j
    \end{pmatrix} + \gv\epsilon
    \label{eq:state_to_obs}
\end{equation}
\noindent where $\gv\epsilon$ is the observational error.

The infected compartments are active variables which can increase and decrease over time because a fraction of the population will transit from $I_j$ to $M_j$, $T_j$ or $C_j$, and then will be accumulated in $R_j$ or $D_j$, that is why these variables need to be considered for the accumulated cases. 

For the parameter estimation, the model parameters $\gv \theta$ and the state variables $S_j$, $E_j$, $I_j$, $M_j$, $T_j$, $C_j$, $R_j$, $D_j$ are put together in an augmented state vector $\v x$. Then, the model parameters are estimated in the same way as the state variables, using the EnKF. This technique is known as the \textit{augmented state}. A review of parameter estimation using various data assimilation methods based on the state augmentation approach can be found in \cite{ruiz13}. The fractions of detected cases $\gamma_j$ are also estimated in this way. Although these parameters are not part of the model equations, their estimation can be conducted in the same way as for the model parameters.

  The parameters in ensemble-based data assimilation are estimated through their correlations with the observed variables. Therefore, parameter estimation depends crucially on an accurate quantification of the augmented error covariance matrix.  While chaotic dynamics drives the evolution in state variables leading  to an increase in their ensemble spread, persistence is assumed for the time evolution of the parameters. Because of this an inflation method is required to prevent the parameter ensemble spread from collapsing in an ensemble data assimilation cycle (e.g. Ruiz et al. 2013) \cite{ruiz13}.

  We performed preliminary simulations to evaluate the use of multiplicative inflation in the EnKF framework. Even when we use two independent inflation factors, one for the parameters and one for the state variables \cite{ruiz13b}, it was not posible to find a suitable set of inflation factors, they resulted in filter divergence or poor estimation performance. Because of this, we opted for another approach: to model the parameter evolution of each ensemble member as an independent auto-regressive process or correlated random walk \cite{liu01} with correlation $\rho$ and standard deviation $\sigma$, which is applied only to the estimated parameters $\v \theta^i$ (no inflation is applied to the state variables):

\begin{equation}
    \gv\theta^i_{k+1} = \bar{\gv\theta}_k +\rho~ (\theta^i_{k}-\bar{\gv\theta}^i_{k}) + \sigma \sqrt{1-\rho^2}\, \v \eta (0,1)\label{rWalk}
\end{equation}
\noindent where $\eta\left(0,1\right)$ is a random Gaussian number with zero mean and unitary standard deviation. The inflation is added before the analysis step.

To summarize our estimation method, the EnKF methodology is represented concisely in Algorithm 1.

\begin{algorithm}[H]
    \caption{Stochastic ensemble Kalman Filter}\label{alg:enkf}
    \begin{algorithmic}
        \Require{$\v H$, $\v R$, $\mathcal M()$ and $\textbf{x}^{\text{a},(i)} = \textbf{x}^{\text{f},(i)}(t=t_0)$, i=1,\,...\,m} \Comment{Inputs and ensemble initialization}
        \State \textbf{do} $t_k=1,2,...$
        \State \indent $\textbf{x}^{\text{f},(i)} = \mathcal M(\textbf{x}^{\text{a},(i)})$ 
        \State \indent $\textbf{P}^\text{f} = \frac{1}{m-1}\sum^m_{i=1}(\textbf{x}^{\text{f},(i)} - \overline{\textbf{x}}^\text{f})(\textbf{x}^{\text{f},(i)} - \overline{\textbf{x}}^\text{f})^\top$  \Comment{forecast covariance}
        \State \indent $\textbf{K} = \textbf{P}^\text{f}\textbf{H}^\top (\textbf{H} \textbf{P}^f \textbf{H}^\top + \textbf{R})^{-1}$ \Comment{Kalman gain}
        \State \indent $\textbf{y}^{(i)} = \textbf{y} + \gv\mu^{(i)},$ \Comment{Perturbed observations}
        \State \indent $\gv\theta^{(i)}_{t_k} = \bar{\gv\theta}_k +\rho~ (\gv\theta^i_{k}-\bar{\gv\theta}^i_{k}) + \sigma \sqrt{1-\rho^2}\, \v \eta (0,1)$ \Comment{Inflate parameters}
        \State \indent $\textbf{x}^{\text{a},(i)} = \textbf{x}^{\text{f},(i)} + \textbf{K}(\textbf{y}^{(i)} - \v H\textbf{x}^{\text{f},(i)})$ \Comment{Analysis}
        \State \textbf{end do}
    \end{algorithmic}
\end{algorithm}

\section{Experimental details} \label{detalles_exp}

\subsection{Transmission matrix parameterizations} \label{cmat_par}

For an $n \times n$ transmission matrix there are $\frac{n^2+n}{2}$ independent parameters to be estimated instead of $n^2$ because of the restriction \reff{eq:trm_constraint}. In our case we use three agegroups, so the resulting transmission matrix is

\begin{equation}
    \Lambda =
    \begin{pmatrix}
    \lambda_{11}                 &       \lambda_{12}                 & \lambda_{13} \\
    \frac{N_1}{N_2} \lambda_{12} &       \lambda_{22}                 & \lambda_{23} \\
    \frac{N_1}{N_3} \lambda_{13} &    \frac{N_2}{N_3}\lambda_{23}     & \lambda_{33}
    \end{pmatrix}
    \label{eq:7parmat}
\end{equation}
where parameters $\lambda_{ij}$ depend on time.

As is shown in the experiments in Section \ref{resultados}, the  parameters of \reff{eq:7parmat} are not identifiable if only information of the accumulated infection cases in each group is available, without information about which agegroup was the cause of the new exposed.

To overcome this limitation, we propose a parameterization for the  transmission matrix with fewer parameters:
\begin{equation}
    \Lambda =
    \begin{pmatrix}
    \lambda_{1}                                             &       \alpha \sqrt{\lambda_{1}\lambda_{2}}                    & \alpha \sqrt{\lambda_{1}\lambda_{3}} \\
    \frac{N_1}{N_2} \alpha \sqrt{\lambda_{2}\lambda_{1}}    &       \lambda_{2}                                             & \alpha \sqrt{\lambda_{2}\lambda_{3}} \\
    \frac{N_1}{N_3} \alpha \sqrt{\lambda_{3}\lambda_{1}}    &     \frac{N_2}{N_3}\alpha\sqrt{\lambda_{3}\lambda_{2}}        & \lambda_{3}
    \end{pmatrix}
    \label{eq:3parmat}
\end{equation}
from now on, we call this matrix the parameterized transmission matrix.

This parameterization is a particular case of \reff{eq:7parmat} where the upper diagonal parameters $ij$ are defined as a function of the diagonal elements of the row $i$ and column $j$: $\lambda_{ij} = \sqrt{\lambda_{i}\lambda_{j}}$, and the lower diagonal parameters are defined by the constrain \reff{eq:trm_constraint}. The parameter $\alpha$ controls the relative importance of inter-agegroup and intra-agegroup infections, with lower values giving more weight to the later.

\subsection{Data} \label{data}

We use three agegroups in the range of $[0,30)$, $[30,65)$ and $[65,111]$ years. This division is motivated because we want to represent agegroups with different activities, so that children and young individuals activities are mainly school and universities, adults is the working agegroup and the senior population assumed to be mainly retired. At the same time, these groups grossly represent different health profiles, with senior population the ones that most likely will develop severe symptoms, while the first age group are expected to have minor symptoms. The total population is assumed to be 44.8 millions divided in the three age groups by $2.2\times 10^7$, $1.8\times 10^7$ and $4.8\times 10^6$, which represent the approximate number of people within the aforementioned agegroups in Argentina (taken  from last population census).

\subsubsection{Synthetic observations}\label{data_synt}

The synthetic observations are generated evolving a meta-population model with a prescribed "true" transmission matrix.
Cumulative infected cases and deaths disaggregated by age groups are assumed to be daily observed during a period of 300 days.

The model uses a transmission matrix which has the form \reff{eq:7parmat}, and the parameters $\lambda_{ij}$ are defined as

\begin{equation}
    \gv \lambda =
    \begin{cases}
        [1.6, 1.8, 1.4, 0.5, 0.4, 0.3] & \text{if}\,\, t \in [0,80)\, \ud  \\
        [0.4, 0.6, 0.3, 0.15, 0.13, 0.1] & \text{if}\,\, t \in [80,140)\,  \ud   \\
        [1.6, 1.2, 1.35, 0.36, 0.25, 0.2] & \text{if}\,\, t \in [140,300]\, \ud
    \end{cases}
\end{equation}

\noindent where $\gv \lambda = [\lambda_{11},\lambda_{22},\lambda_{33},\lambda_{12},\lambda_{13},\lambda_{23}]$.

The decrease in the transmission matrix parameters at $t=80$ mimics the effect of a lockdown. Then, the increase at time 140 represents a relaxation to normal conditions but with some sanitary measures (e.g. social distancing, mandatory use of masks in public spaces, etc). These conditions result in a double outbreak situation as observed in Argentina (and several other countries) in the first year of the pandemic.

Note that the relative changes in the parameters are different for different agegroups (i.e. not proportional). We chose on purpose a transmission matrix that cannot be fully represented by the parameterization \reff{eq:3parmat}, so that the model used in the estimation is not perfect (some structural uncertainty is introduced in the parameterization process). Another motivation was to represent the resulting different levels of mobility that were found in different agegroups.

The true values of the fraction of detected cases $\gamma_j,\, j=1,2,3$ are taken to be $0.15$, $0.2$ and $0.3$ corresponding to the young, adult and senior agegroups. A reference single population detection fraction was estimated in \cite{li20}. Intuitively, we expect a higher fraction of symptomatic for the elder agegroups, as it is the most vulnerable population. The fraction of deaths $f^C_j$ of each agegroup is assumed to be $0.002$, $0.05$ and $0.1$ \cite{fd}.

Synthetic observations are generated taking daily values from the true system evolution and adding observational error realization from zero mean Gaussian noise with standard deviation proportional to true value up to a maximum value. After some preliminary experiments we set the standard deviation of the accumulated cases observational error to $\text{\text{max}(0.05 $y^c_j$, 100)}$, where $y^c_j$ indicates the observed cumulative cases for every agegroup j. We assume that deaths are well documented so the standard deviation of the deaths observational error is $\text{\text{min}(0.05 $y^c_j$, 5)}$. The way we define the observational error means that eventually all the observations will have the upper limit error variance after some time.

\subsubsection{Real world observations}\label{data_reales}

For the real world experiments we use epidemiological data from Argentina collected by the National Health Surveillance System (SNVS, for its acronym in Spanish). The SNVS dataset is openly available
(\href{http://datos.salud.gob.ar/dataset/covid-19-casos-registrados-en-la-republica-argentina}{http://datos.salud.gob.ar/dataset/covid-19-casos-registrados-en-la-republica-argentina}) and consists in all the reported tests from public and private tests. The available information for each case is, among other data, the date of the test, the province of residence, age, and whether the person required hospitalization, intensive cares and/or respiratory support.

The first case of SARS-CoV-2 in Argentina was reported on March 3, 2020. Just after 16 days  of this on March 19, 2020, a nationwide lockdown was established.

The data used in the real-world assimilation experiments will be daily cumulative cases and deaths aggregated over the selected agegroups.

\section{Results} \label{resultados}

We present our results in the following order:
\begin{itemize}
    \item In the subsection \ref{exp_sinteticos} we evaluate the model and data assimilation framework with twin experiments.
    \item In the subsection \ref{exp_reales} we apply the methodology to COVID-19 data of Argentina.
    \item In the subsection \ref{pronosticos} we conduct forecasts to examine the performance of the meta-population model coupled with the EnKF using the real observations.
\end{itemize}

\subsection{Experiments with synthetic observations} \label{exp_sinteticos}

The objective of the twin experiments is to evaluate the data assimilation-based parameter estimation in a context in which the true parameters are known and errors in the estimation can be accurately computed.

The assimilation filter estimates all the variables of the system and the parameters of the transmission matrix, which are augmented to the system state vector.The dimension of the augmented state vector is 24 and the amount of estimated parameters is six in the case of the parameterized transmission matrix: three belonging to the parameterized transmission matrix and three corresponding to the fractions of detected cases. In the case of the transmission matrix \reff{eq:7parmat}, there are nine estimated parameters: six from the matrix and three from the fraction of detected cases so that the augmented state vector dimension is 27.

As mentioned, the EnKF for parameter estimations requires an inflation approach for the parameter spread \cite{ruiz13}. 
The filter was able to track the observations using the correlated random walk, \reff{rWalk}, for high values of $\rho (0.999)$ and $\sigma$ in the range $[0.001,0.2]$. We measured the RMSE of the estimation compared to the true values of the cases, of the deaths and of the transmission matrix parameters. Each RMSE showed a different optimal value of $\sigma$. We took $\sigma=0.05$ and $r=0.999$ which results in almost optimal estimates of the parameterized transmission matrix parameters and at the same time having good estimations of the state variables. The same random walk parameter values are used in the real data experiments (section \ref{exp_reales}).  

\begin{figure}[H]
    \centering
    \includegraphics[width=.7\textwidth]{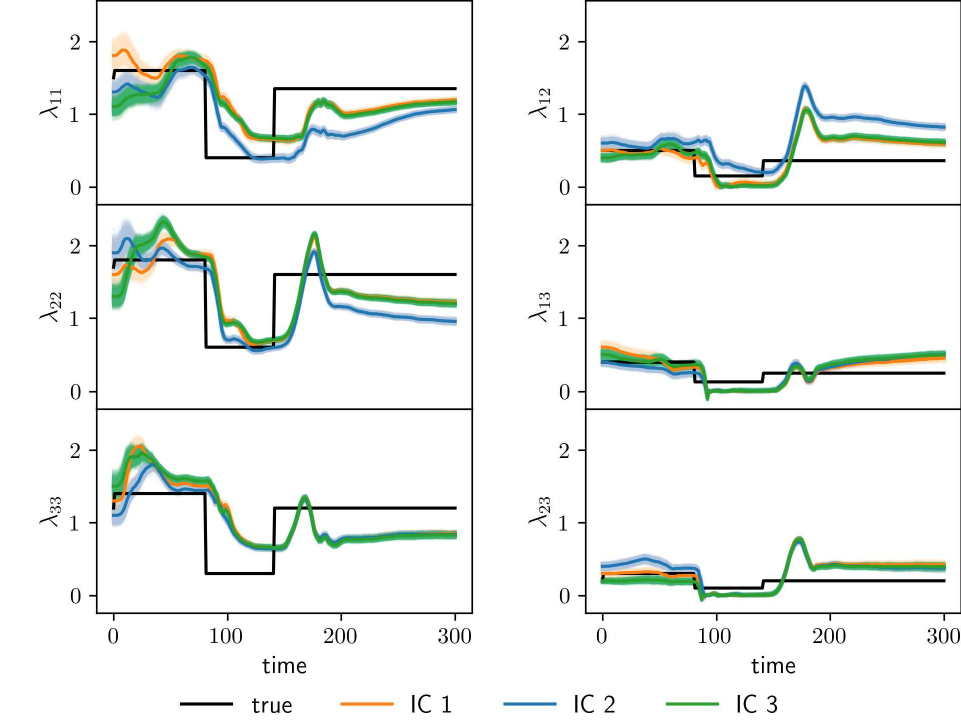}
    \caption{\footnotesize Estimated parameters of the full transmission matrix. Left panels: diagonal parameters. Right panels: off-diagonal parameters. Colored curves represent estimations with different initial conditions, and black curves represent the true parameter values. Shades around colored curves represent the parameter spread.}
    \label{fig:7par_cmat}
\end{figure}

Fig \ref{fig:7par_cmat} shows the estimated parameters for the twin experiments using the six-parameter transmission matrix \reff{eq:7parmat}. To examine the identifiability and sensitivity to initial conditions of the estimated parameters, three experiments with different apriori density of parameters at $t=0$ are shown. Some of the time variability of the true parameters is captured, however the different experiments converge to different estimated parameter values. The estimations of the parameters are dependent of the initial condition in the sense that different initial conditions of the parameters result in different estimations of the parameters at later times ($> 100~\text{d}$), and neither of the three experiments is able to estimate precisely the true parameters (Fig \ref{fig:7par_cmat}).
The reason for this is that an increase in the rate of cases, say in the agegroup 1, may be ascribed by the assimilation system to a change in the parameters $\lambda_{12}$ or to a change in $\lambda_{11}$ and $\lambda_{22}$. Both scenarios result in the same infection rates so that the information provided by the observations is not enough to identify the actual scenario. For instance, $\lambda_{33}$ green curves in Fig  \ref{fig:7par_cmat} present an underestimation at the beginning of the lockdown, this underestimation is balanced with the overestimation of $\lambda_{13}$ and $\lambda_{23}$, leading to an evolution of the number of cases consistent with the observations. We point out that the estimation of observed variables, the cases and deaths, is equally accurate for all these experiments. For this reason we conclude that the six parameter transmission matrix is not identifiable using age-dependent cases and deaths observations data.

 There is some delay in the estimated transmission matrix parameters shown in Fig \ref{fig:7par_cmat} between the abrupt change due to the lockdown measure (both in the beginning and end) that we imposed to the true parameter and the captured change in the estimated parameter. Estimated parameters start to adjust to these abrupt changes a few days after the change and they converge to a new value 20-30 days after. The reason for this is that  parameters in ensemble based assimilation systems are estimated through the correlation with observed variables, so that these state-parameters correlations take some cycles to adapt to abrupt changes. This behavior can be reduced by tuning up the amount of inflation, at the expense of having an increased spread  in the estimated parameter and state variables ensemble. Overall, the amplitude of the abrupt change is rather well estimated beyond the mentioned delay.

\begin{figure}[H]
    \centering
    \includegraphics[width=.9\textwidth]{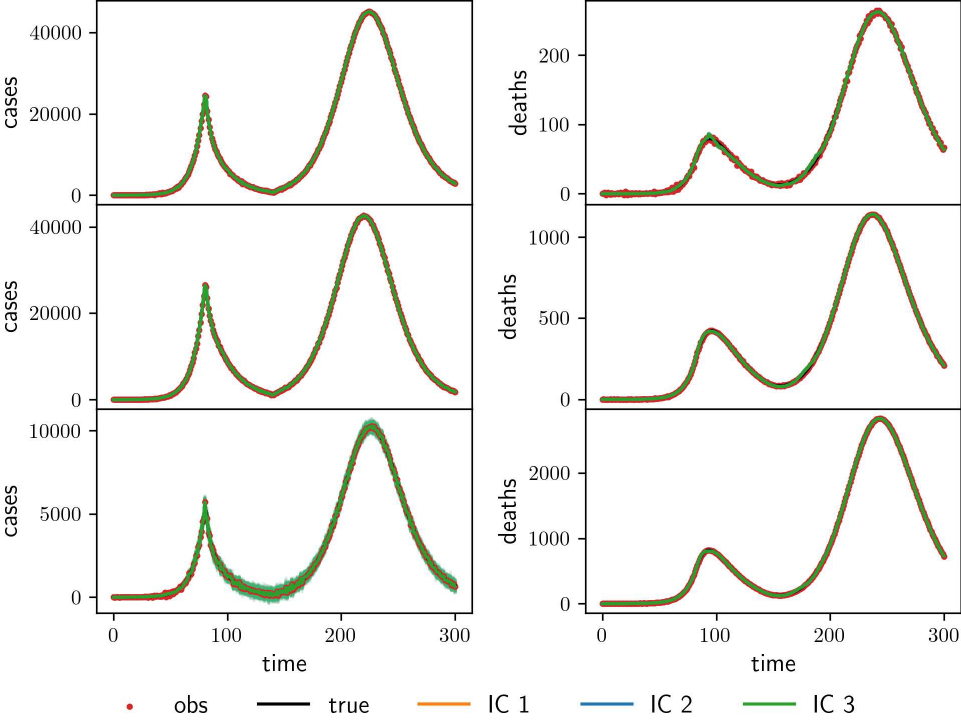}
    \caption{\footnotesize Estimated incident cases (left panels) and deaths (right panels) of the young (upper panels), adult (middle panels) and senior (lower panels) agegroups for the full transmission matrix experiment. Colored curves represent estimations with different initial conditions, red dots represent observations and black curves represent the true parameter values. Shades around colored curves represent the corresponding variable spread.}
    \label{fig:3par_cases_deaths}
\end{figure}

Fig \ref{fig:3par_cases_deaths} shows the estimated daily new cases (left panels) and deaths (right panels) of the young (upper panels), adult (middle panels) and senior (lower panels) agegroups, using the full transmission matrix \ref{eq:7parmat}.The three similar experiments with different initial conditions of the estimated parameters give similar results (curves of the three experiments are indistinguishable in Fig \ref{fig:3par_cases_deaths}). In the three experiments, the EnKF is able to keep track of the observations of cases and deaths in all the agegroups, even though the transmission matrix parameters are not identifiable. The ensemble dispersion in the senior agegroup is relatively larger because the population is almost five times lower than in the other agegroups, and all the agegroups have the same observation error upper limit, so that the relative error of the estimation is higher.

\begin{figure}[H]
    \centering
    \includegraphics[width=.8\textwidth]{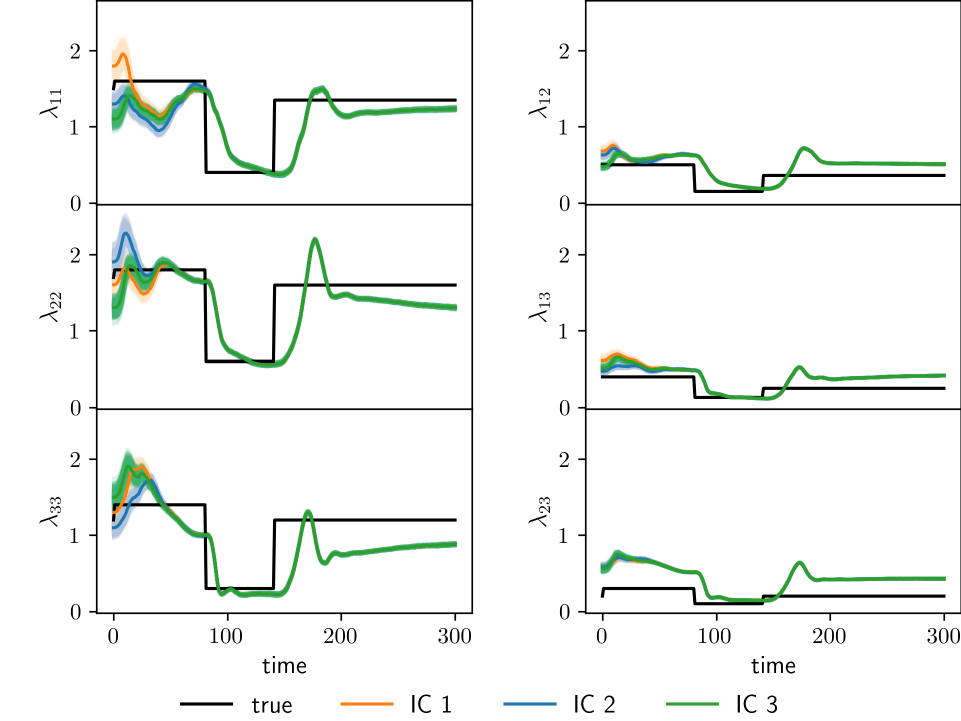}
    \caption{\footnotesize Estimated parameters of the parameterized transmission matrix. Left panels: diagonal parameters. Right panels: off-diagonal parameters. Colored curves represent estimations with different initial conditions, and black curves represent the true parameter values. Shades around colored curves represent the parameter ensemble.}
    \label{fig:3par_cmat}
\end{figure}

Given that the transmission matrix parameters are not identifiable  using the matrix form \reff{eq:7parmat}, we conduct estimation experiments using the proposed parameterization \reff{eq:3parmat}. We took $\alpha=0.4$ in \reff{eq:7parmat}, which represent significant intra-group contagions.
 Fig \ref{fig:3par_cmat} shows the estimated parameters of the parameterized transmission matrix. The right panels show the values of the diagonal, and the left ones show the values of the upper off-diagonal.

The three experiments converge to the same parameter values, independently of the initial condition. The true values of the parameters cannot be estimated precisely because this parameterization is not able to exactly fit the structure of the true transmission matrix. Because of this, parameters representing intra-group contacts are systematically underestimated while the number of  inter-group interactions are overestimated. Note that this bias could be alleviated with a lower $\alpha$ value, however this optimization based on true transmission matrix values cannot be conducted in realistic cases. The parameter estimates in Fig \ref{fig:3par_cmat} also show a delay in the representation of the sudden parameter changes found at the beginning and at the end of the lockdown period, as found in Fig \ref{fig:7par_cmat}.

\begin{figure}[H]
    \centering
    \includegraphics[width=.8\textwidth]{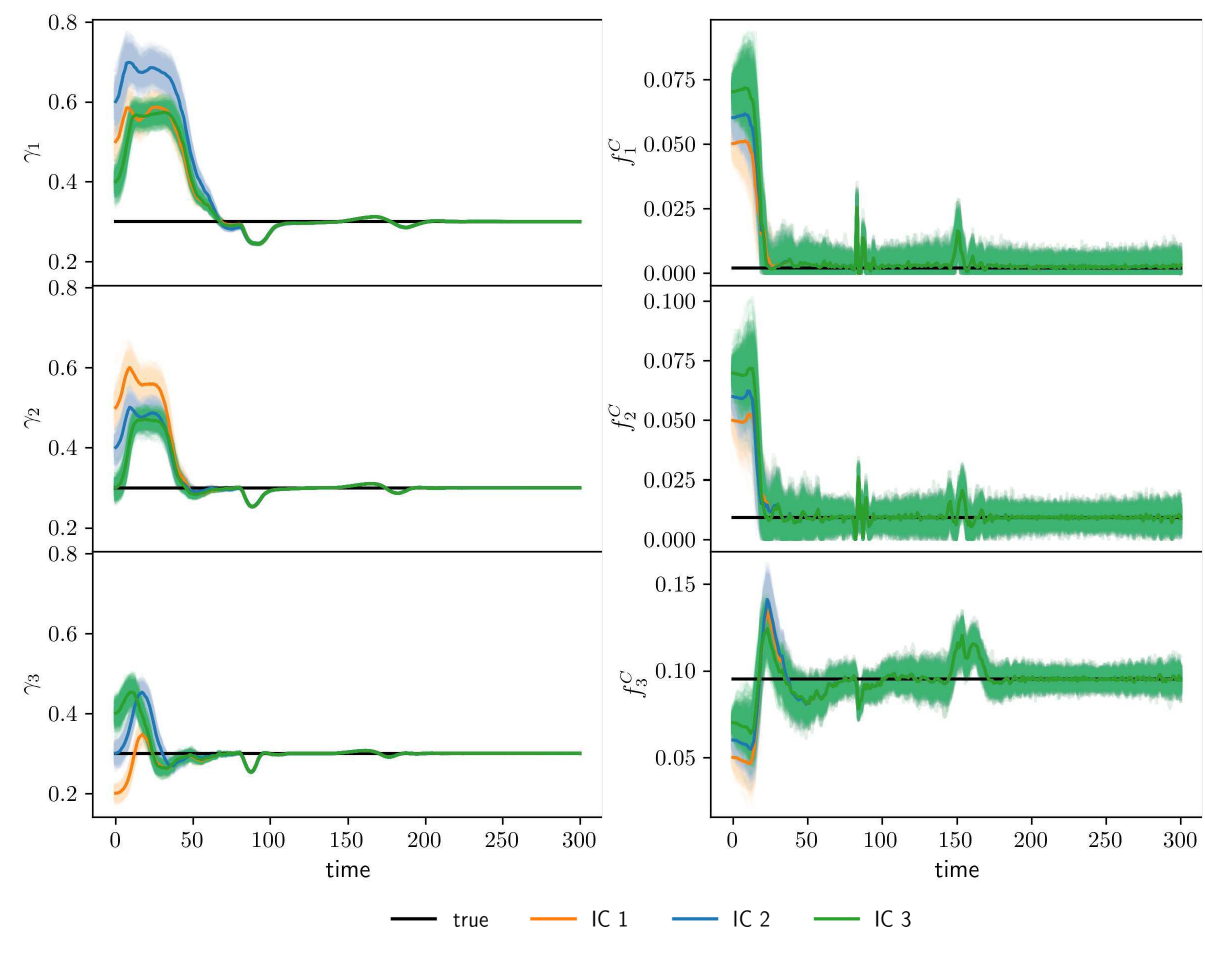}
    \caption{\footnotesize Left: estimated fraction of detected cases of each agegroup. Right: estimated fraction of deaths of each agegroup. Both group of plots correspond to different simulations. Colored curves represent estimations with different initial conditions, and black curves represent the true parameter values. Shades around colored curves represent the parameter ensemble.}
    \label{fig:3par_fa_fd}
\end{figure}

Fig \ref{fig:3par_fa_fd} shows the fraction of detected cases of each agegroup (left panels). We expect these parameters to be correlated to observed accumulated cases and deaths. Therefore, the system should be able to constrain them. The true values of $\gamma_j$ are accurately estimated by the assimilation system, regardless of the initial condition. The spurious peaks estimated in the parameterized transmission matrix at the lockdown transitions are also found in the $\gamma_j$ parameters around time 80 and, with much less intensity, at 170.

In the previous shown experiment, we estimated a parameterized transmission matrix and the fraction of detected cases $\gamma_j$. The cases, deaths and the parameters $\lambda_{j}$ can also be estimated alongside with the fractions of deaths $f_j^C$ instead of $\gamma_j$. To illustrate this, we fix $\gamma_j$ equal to the true values and perform three experiments that estimate the transmission matrix and the fraction of deaths. The parameters $\lambda_j$ are similar to the ones showed in Fig \ref{fig:3par_cmat}. The obtained $f_j^C$ estimates are shown in the right panels of Fig \ref{fig:3par_fa_fd}. In all experiments the estimated parameters converge to the true values, and the sudden change in the estimations is again observed at the times where the true transmission matrix parameters change. 

 In figure \ref{fig:3par_fa_fd} we can see the lower bound in the estimated parameter $f^C_1$: the true value is near zero, and statistically some of the filter corrections tend to be negative, which are then corrected (all the values are positive).

\begin{figure}[H]
    \centering
    \includegraphics[width=1.\textwidth]{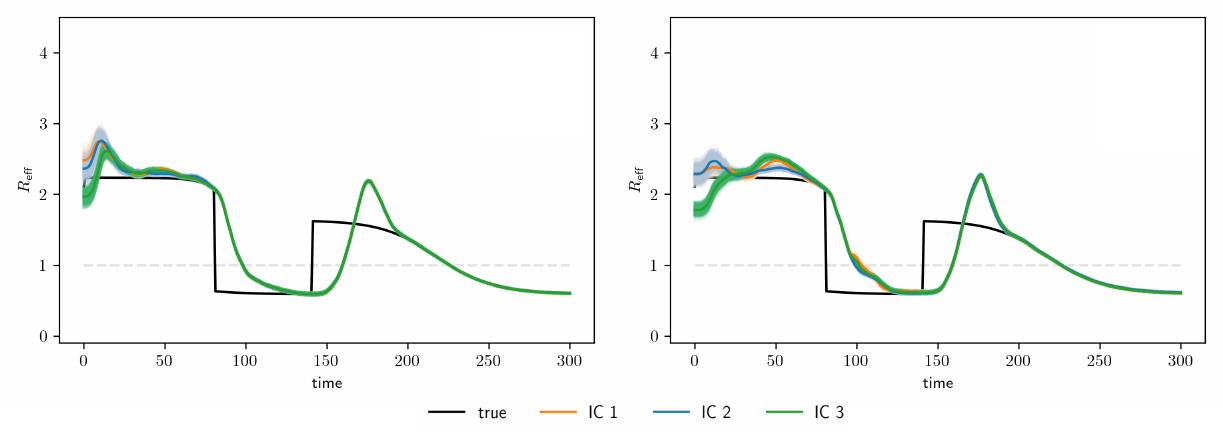}
    \caption{\footnotesize Estimated effective reproduction number using the parameterized transmission matrix (left) and the six-parameter transmission matrix (right). Both group of plots correspond to different simulations. Colored curves represent estimations with different initial conditions, and black curves represent the true parameter values. Shades around colored curves represent the parameter ensemble.}
    \label{fig:3par_Reff}
\end{figure}

Fig \ref{fig:3par_Reff} shows the effective reproduction number computed with the next generation matrix for the experiment corresponding to the  parameterized transmission matrix (left panel) and to the six-parameter matrix (right panel). In both cases, the true values of $R_{\text{eff}}$ can be accurately estimated with both parameterizations (apart from the delay in parameter changes), even when the true transmission matrix is non-reproducible by the parameterized transmission matrix. This result can be interpreted as follows: Our parameterized transmission matrix is flexible enough to capture the system $R_{\text{eff}}$ and its temporal evolution and at the same time have a dimensionality low enough to allow its parameters to be identifiable from the available observations.

\subsection{Real data experiments} \label{exp_reales}

An experiment is conducted with the same  assimilation system as in the previous section  using the real dataset of Argentina. Contrary to the twin experiments, the observations may be biased and the observational error covariance is unknown. Indeed, the observed cases are highly noisy. One of the sources of the noise is due to the fact that testing and report diminish on weekends, resulting in an under-report of cases during weekends and likely an over-report on Mondays and Tuesdays due to delayed reports.

The observations consist on accumulated cases and deaths for each agegroup in the time interval from 2020/03/03 to 2021/09/18 (564 days).
We estimate the parameterized transmission matrix \ref{eq:3parmat} with $\alpha=0.5$ and the time-dependent fraction of deaths is estimated in the real observation experiments to account for the following effects: the data correspond to a time interval of almost 1.5 years, time in which the SARS-Cov-2 virus mutated several times changing the severity of the symptoms; and two, because of improvements of symptom treatments in the health system and the start of the vaccination period. 

The fraction of detected cases were assumed to be fixed at $0.2$, $0.3$ and $0.4$. The observation error is set to $\text{\text{max}(0.05 $y^c_1$, 400)}$, $\text{\text{max}(0.05 $y^c_2$, 500)}$, $\text{\text{max}(0.05 $y^c_3$, 50)}$ in the young, adult and senior agegroups respectively, and $\text{\text{max}(0.05 $y^d_i$, 5)}$ for the death observations in all agegroups.

\begin{figure}[H]
    \centering
    \includegraphics[width=.9\textwidth]{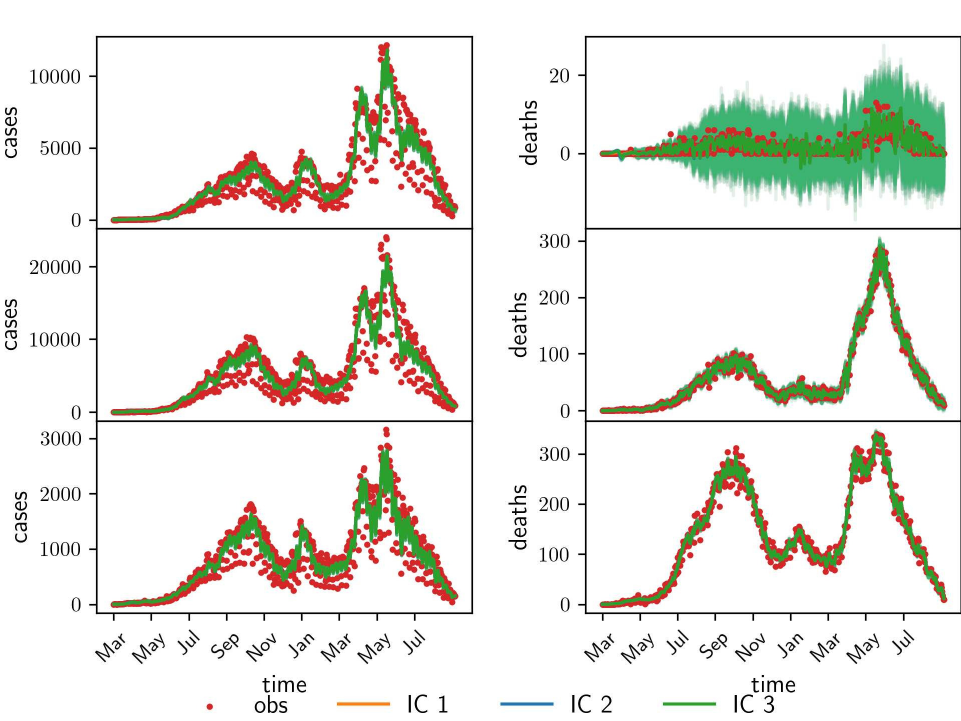}
    \caption{\footnotesize Estimated incident cases (left-side panels) and deaths (right-side panels) in the real data experiments using the parameterized transmission matrix. Young agegroup: upper panels. Adults agegroup: middle panels. Senior agegroup: bottom panels. Colored curves represent estimations with different initial conditions and red dots represent observations. Shades around colored curves represent the estimated variable uncertainty.}
    \label{fig:3par_cases_deaths_real_data}
\end{figure}

Fig \ref{fig:3par_cases_deaths_real_data} shows the incident cases (left panels) and incident deaths (right panels) of the young (top panels), adult (middle panels) and senior (bottom panels) agegroups, respectively. The filter is able to keep track of the observations of each agegroup since the cases and deaths are estimated correctly. As in the twin experiments, we use three sets of initial conditions, they yield the same estimation of cases and deaths (in Fig \ref{fig:3par_cases_deaths_real_data} only the green one is visible). The high frequency cycle found in the estimations of cases and deaths correspond to the weekly observations cycle. If required, this effect can  be mitigated by increasing the observational error of the cases, at the expense of an increase of the uncertainty of the estimations.

The estimated deaths in the young agegroup have a high dispersion because of the relatively few observed cases compared to the other agegroups and their relative errors.
Although the estimated accumulated number of deaths is always positive, the daily changes in the number of deaths is sometimes negative for some ensemble members for the young compartment. This non-physical behavior is a consequence of the updates introduced by the observations which may eventually result in the reduction of estimated number of deaths in order to better fit the observed values.

\begin{figure}[H]
    \centering
    \includegraphics[width=.9\textwidth]{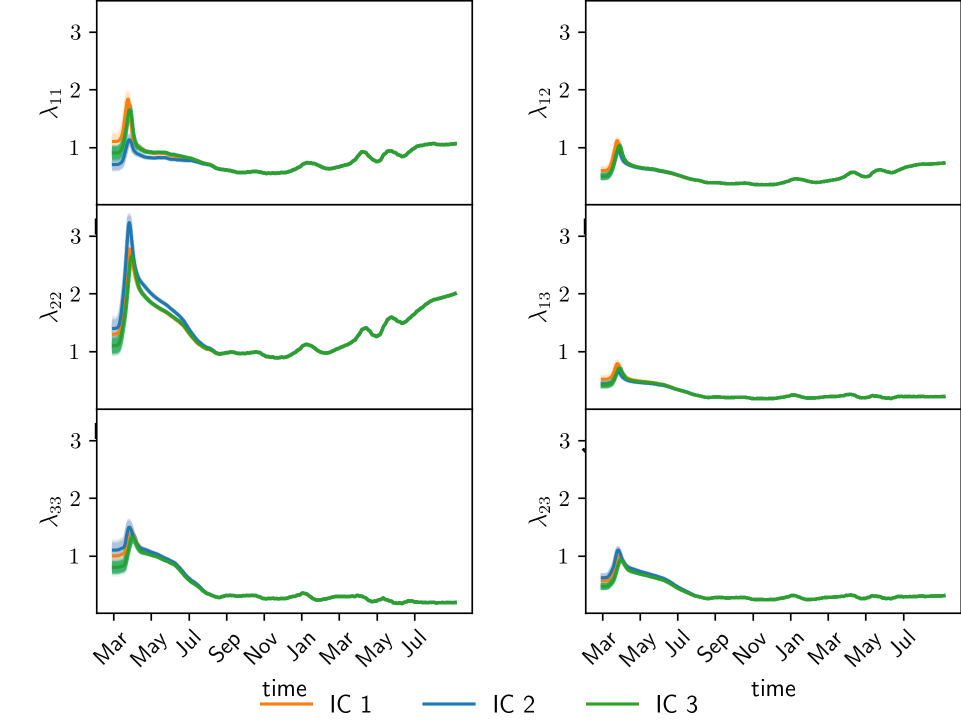}
    \caption{\footnotesize Estimated parameters of the parameterized transmission matrix. Left panels: diagonal parameters. Right panels: off-diagonal parameters. Colored curves represent estimations with different initial conditions, and shades around colored curves represent the parameter spread.}
    \label{fig:cmat_datos_reales}
\end{figure}

Fig \ref{fig:cmat_datos_reales} shows the three independent parameters $\lambda_{i}$, $i=1,2,3$ of the parameterized transmission matrix (left panels), and the upper off-diagonal parameters $\lambda_{ij}=\alpha\sqrt{\lambda_{i}\lambda_{j}}$ (right panels). All different initial conditions yield the same estimations of the parameters. There is a predominance of the parameters $\lambda_{1}$ and $\lambda_{2}$ given that the majority of the cases occur at the first two agegroups. Consequently the interaction parameter between young and adults $\lambda_{12}$ is higher compared to $\lambda_{13}$ and $\lambda_{23}$.

\begin{figure}[H]
    \centering
    \includegraphics[width=.5\textwidth]{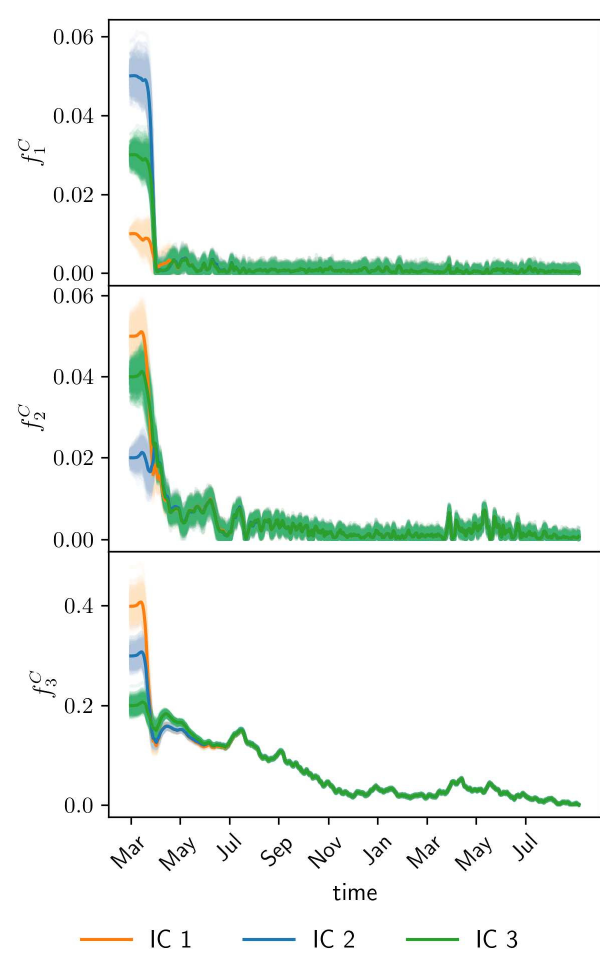}
    \caption{\footnotesize Estimated fraction of deaths of each agegroup. Colored curves represent estimations with different initial conditions, and shades around colored curves represent the corresponding variable spread.}
    \label{fig:fd_datos_reales}
\end{figure}

Fig \ref{fig:fd_datos_reales} shows the fraction of deaths of each agegroup. Once more we can see the independence of the estimation over the initial condition. The estimated parameter of the young population presents a high ensemble dispersion because of the few deaths observed in the agegroup.

The estimated fractions of deaths are much higher than the reference values $0.002$, $0.05$ and $0.1$ \cite{fd} of the young, adult and senior agegroups. This discrepancy may be a consequence of the under-detection of cases: the fraction of deaths need to rise for the system to make sense of the lack of cases. Also the lower bound of the estimated parameters may contribute to this effect.

\begin{figure}[H]
    \centering
    \includegraphics[width=.9\textwidth]{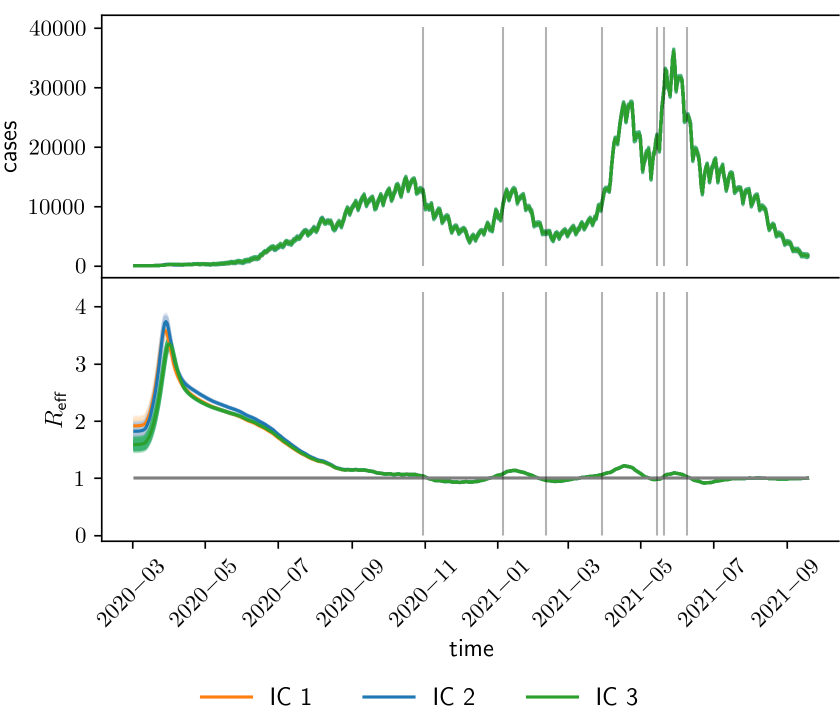}
    \caption{\footnotesize Upper panel: total incident cases in Argentina. Lower panel: estimated reproduction number using COVID-19 data of Argentina. Different colors indicates different initial conditions, and shades around colored curves represent the corresponding variable spread. Vertical lines point out periods where $R_{\text{eff}}>1$.}
    \label{fig:Reff_datos_reales}
\end{figure}

Fig \ref{fig:Reff_datos_reales} shows the estimated effective reproduction number $R_{\text{eff}}$. The estimated parameter does not depend on the chosen initial conditions. The estimated periods where $R_{\text{eff}}>1$ corresponds to an increase in the cases up to the peaks.

\subsection{Forecasts}   \label{pronosticos}

\begin{figure}[H]
    \centering
    \includegraphics[width=.8\textwidth]{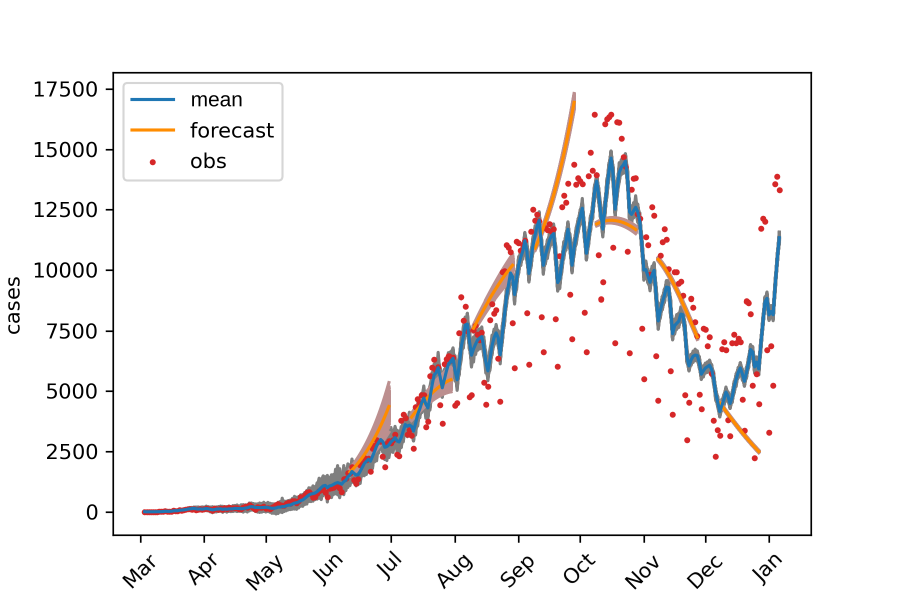}
    \caption{\footnotesize Forecasts (orange curves) conducted in different stages of the pandemic in Argentina, observed cumulative cases (showed as incidence/daily cases; red dots) and analysis of cases (blue curves) using the EnKF with multiplicative inflation. Shades around curves represent ensemble members.}
    \label{fig:forecasts}
\end{figure}

To evaluate the potential use of the estimated parameters for decision making, we conducted an evaluation of the performance of the resulting forecasts using the estimated parameterized transmission matrix on the COVID-19 data of Argentina. The methodology is as follows. First linear and quadratic fits are performed on the last 15 days of the estimated parameterized transmission matrix values to obtain the parameter tendencies. Then, these tendencies are projected 30 days forward, starting from the last value of the analysis (current day). Finally, the 30-day forecasts are conducted with the free evolution of the model using the projected parameterized transmission matrix and starting from the current analysis state. We compare the forecasts to the assimilation analysis using the entire set of observations over time as true. Fig \ref{fig:forecasts} shows some 30-day forecasts performed over different times of the pandemic. We can see some forecasts are accurate but some diverge from what actually happened (during the tendency changes).

To evaluate the performance of the forecasts, we compute the root mean square error of the incident cases forecast over $400$ different forecasts separated over a time window of $1$ day for each agegroup. To investigate the impact of considering the interactions among different agegroups into the system, we repeat the forecast experiments using a SEIR model with no agegroup division. The relative initial infected of the meta-population model is used as the fraction of the population to obtain the well-mixed model predictions for each agegroup. The initial condition of the well-mixed forecast is the sum along the agegroups of the meta-population model. The forecasts cover the time window from July 2020 to the end of August 2022, featuring two peaks of the infection so that there is a wide variety of epidemic behaviors, as shown in Fig $\ref{fig:forecasts}$.

\begin{figure}[H]
    \centering
    \includegraphics[width=.5\textwidth]{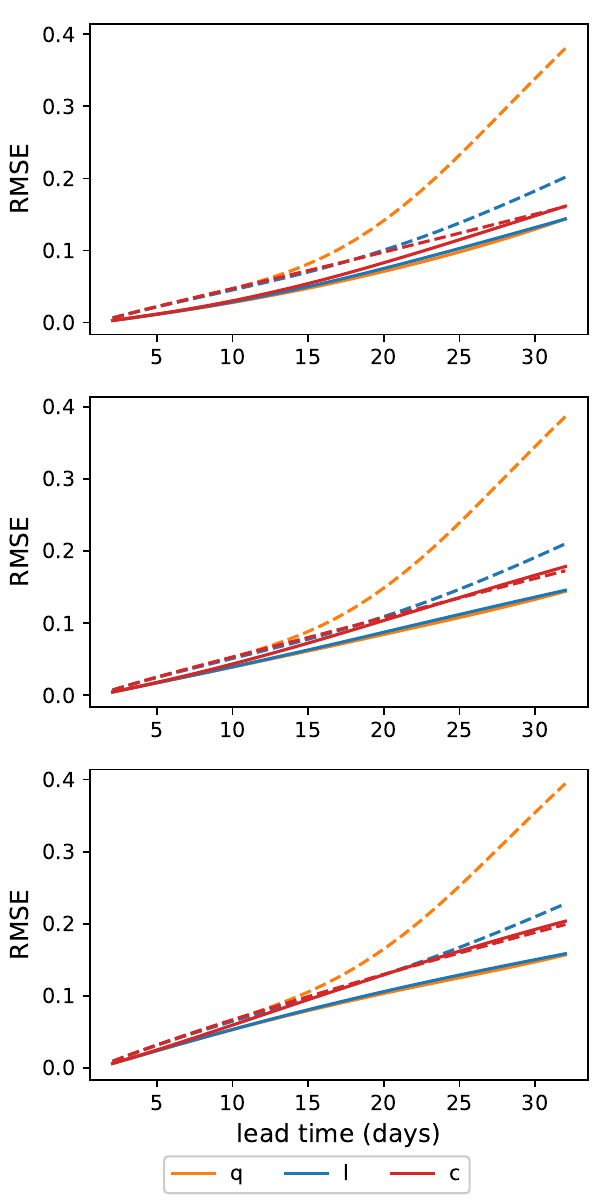}
    \caption{\footnotesize Relative average root mean square error of the quadratic (orange), linear (blue) and constant (red) forecasts using the parameterized transmission matrix (solid plots) and the well-mixed model (dashed plots). Upper panel: young people agegroup. Middle panel: adult people agegroup. Lower panel: senior people agegroup.}
    \label{fig:rmse_forecasts}
\end{figure}

Fig \ref{fig:rmse_forecasts} shows the relative RMSE as a function of the lead time. The behaviour of the forecasts is similar along the agegroups. At the first 15 days the forecasts are similar, but from day 16 to 30 there is an advantage of the quadratic and linear meta-population forecast, closely followed by the constant well-mixed forecast, except for the senior agegroup where the constant well-mixed forecast performs slightly worse than the other  two. The constant meta-population and linear well-mixed forecast shows similar performance, while the less accurate by far is the quadratic well-mixed.

\section{Conclusions} \label{conclusiones}

In this work we used an ensemble Kalman filter applied to a meta-population compartmental model to monitor epidemiological parameters of the SARS-CoV-2 virus and to conduct forecasts. We sequentially calibrated the parameters of the model using state augmentation strategies. Crucially, unlike recent works which use a constant transmission matrix parameters or work directly with well-mixed models, we provided a time-dependent parameterization of the transmission matrix that was identifiable by the system. Besides, in the context of data assimilation, it allows us to detect nontrivial parameter variations and interactions between agegroups which could not be modeled assuming a time independent transmission matrix. Furthermore, other important epidemiological parameters were recovered such as the mortality, fraction of undocumented cases and the effective reproduction number, the last one diagnosed using the NGO. The assimilation technique can be used as a tool for the monitoring and prediction of current and future contagious diseases. We validated the technique with synthetic and real accumulated cases and deaths observations in Argentina.

Three agegroups are used but the technique can be applied to more agegroups containing narrower age ranges for a more precise analysis. Attempting to estimate the full transmission matrix results in the non-identifiability of the parameters. To solve this problem we introduced a parameterization of the transmission matrix \reff{eq:3parmat} because the number of parameters grows linearly with the number of observations. This parameterization introduces a single inter-group transmission parameter, $\alpha$ parameter, which in our experiments was fixed but it could in principle be estimated by performing forecasts in a validation data set (past evolution of the pandemic up to the 'current' pandemic day) and minimizing the relative root mean square error as a function of $\alpha$ at an apriori defined forecast lead time.

In the EnKF framework, we assume errors are  Gaussian which may not be appropriate for some model parameters. Because of this, some model parameters has to be forced to remain within their physically meaningful range. Some model parameters (parameterized transmission matrix, fraction of detected cases and fraction of deaths) are forced to be non-negative to avoid non-physical evolution of the model. This conflicts with the Gaussian assumption particularly when the spread of the variable or parameter are close to the boundaries of their meaningful range. This is the case for the fraction of deaths in the young population. A non-parametric data assimilation framework can be applied, like the mapping particle filter \cite{pulido19}, to avoid this limitation and to represent the non-Gaussian density of the near-zero parameters.
Furthermore, the variables are assumed to evolve with a smooth behavior, which is achieved for a relatively large number of individuals (country-level observations). In the case of city-level populations, the behavior of the age-meta-population model within the EnKF framework may not be robust, increasing granularity in agegroups and contacts can be achieved by using epidemiological agent-based models. Recently,  Cocucci et al. (2022) \cite{cocucci22} used an EnKF combined with an ABM using mean field data to infer the COVID-19 pandemic in the city of Buenos Aires, Argentina. Schneider et al. (2022) \cite{schneider22} used a complex agent-based network model to assimilate synthetic data at individual level.


 The use of the meta-population model resulted in an improvement of the forecasts up to 30 days lead times of the new cases compared to well-mixed models, which does not account for the interaction of compartments among different agegroups. This highlights the importance of disaggregating information in both data and model. The age dependent forecasts may be of interest  considering epidemiological models were used by governments in the pandemic decision making. 

We evaluated different parameter regression functions for the transmission matrix values which are then extrapolate temporally to conduct the forecasts. Up to 15 day lead times, there is practically no difference in the forecast accurate between the three regression functions (constant, linear quadratic), but for longer lead times, the quadratic and linear regression functions give the extrapolated values which results in the most accurate forecasts.

Our framework could be greatly improved including hospitalizations as an observed variable.
If reliable data of check-in and check-out hospitalizations were available, relevant quantities could be estimated, like average hospitalization times and use of hospital beds, and also parameters like fraction of hospitalizations and fraction of intensive care. 


%

%


%

\end{document}